\documentclass[sigconf,natbib=false,authorversion]{acmart}

\makeatletter
\def\@ACM@checkaffil{
    \if@ACM@instpresent\else
    \ClassWarningNoLine{\@classname}{No institution present for an affiliation}%
    \fi
    \if@ACM@citypresent\else
    \ClassWarningNoLine{\@classname}{No city present for an affiliation}%
    \fi
    \if@ACM@countrypresent\else
        \ClassWarningNoLine{\@classname}{No country present for an affiliation}%
    \fi
}
\makeatother

\AtBeginDocument{%
  }

\usepackage{cite}
\usepackage{amsmath,amsfonts}
\usepackage{algorithmic}
\usepackage{graphicx}
\usepackage{textcomp}
\usepackage{caption}
\usepackage{subcaption}
\usepackage{multirow}
\usepackage{adjustbox}

\usepackage{hyperref}

\newcommand{\iu}{\institution{\small Indiana University}\city{ Bloomington}\state{IN}\country{USA}}
\newcommand{\pnnl}{\institution{\small Pacific Northwest National Laboratory}\city{ Richland}\state{WA}\country{USA}}
\newcommand{\ksu}{\institution{\small Kent State University}\city{Kent}\state{OH}\country{USA}}
\newcommand{\thiswork}{\textsc{MEMQSim}}

\copyrightyear{2023}
\acmYear{2023}
\setcopyright{licensedusgovmixed}
\acmConference[SC-W 2023]{Workshops of
The International Conference on High Performance Computing, Network,
Storage, and Analysis}{November 12--17, 2023}{Denver, CO, USA}
\acmBooktitle{Workshops of The International Conference on High Performance
Computing, Network, Storage, and Analysis (SC-W 2023), November 12--17,
2023, Denver, CO, USA}
\acmPrice{15.00}
\acmDOI{10.1145/3624062.3624257}
\acmISBN{979-8-4007-0785-8/23/11}

\begin{document}

\title{\thiswork: Highly Memory-Efficient and Modularized Quantum State-Vector Simulation}

\author{Boyuan Zhang}
\email{bozhan@iu.edu}
\affiliation{\iu}

\author{Bo Fang}
\email{bo.fang@pnnl.gov}
\affiliation{\pnnl}

\author{Qiang Guan}
\email{qguan@kent.edu}
\affiliation{\ksu}

\author{Ang Li}
\email{ang.li@pnnl.gov}
\affiliation{\pnnl}

\author{Dingwen Tao}
\email{ditao@iu.edu}
\affiliation{\iu}

\maketitle

\section{Introduction}

The field of quantum computing has seen a marked advancement \cite{boixo2018characterizing}. However, despite the substantial potential, present-day quantum computing mechanisms are challenged by considerable environmental noise, and the efficacy of quantum error correction in the Noise-Intermediate-Scale-Quantum (NISQ) era remains limited \cite{preskill2018quantum}. Quantum circuit simulation serves as an essential tool for researchers from a variety of disciplines, offering invaluable benefits to validate the accuracy of quantum algorithms \cite{jones2019quest}.

Nonetheless, the simulation of quantum circuits poses significant challenges, primarily because memory utilization escalates exponentially with the increment of qubit quantity. For instance, the Frontier system, with a memory capacity of 47.3 PB, is only equipped to simulate 51 qubits, while the Summit, with a memory capacity of 2.8 PB, can merely simulate 47 qubits \cite{wu2019full}. In light of the reality that vast-memory systems such as high-performance computing (HPC) or cloud infrastructures are not readily available to the majority of practitioners in the field of quantum computing, the capability to simulate the execution of quantum circuits is considerably restricted by devices' memory capacities.

Cutting-edge quantum state-vector simulators \cite{li2021sv, zhang2022uniq, zhang2021hyquas, hisvsim} have, thus far, not placed an emphasis on minimizing the memory footprint during the simulation process. A prior research endeavor \cite{wu2019full} did incorporate compression into state vector simulation with the aim of expanding the number of qubits accommodated within a restricted memory space. This approach, while promising, still presents unresolved complications that necessitate substantial research attention:
(1) In this study, compression and decompression processes occur with high frequency, thereby constituting a significant portion of the total simulation time.
(2) it does not harness the processing power of GPUs to expedite the simulation process. Despite the potential for supporting a greater number of qubits, the subpar performance renders it impractical for wider application.
(3) Data locality is insufficiently utilized, resulting in low cache hit rates. A more efficient memory access pattern is needed.

\begin{figure}[t]
\vspace{-2mm}
	\includegraphics[width=.95\linewidth]{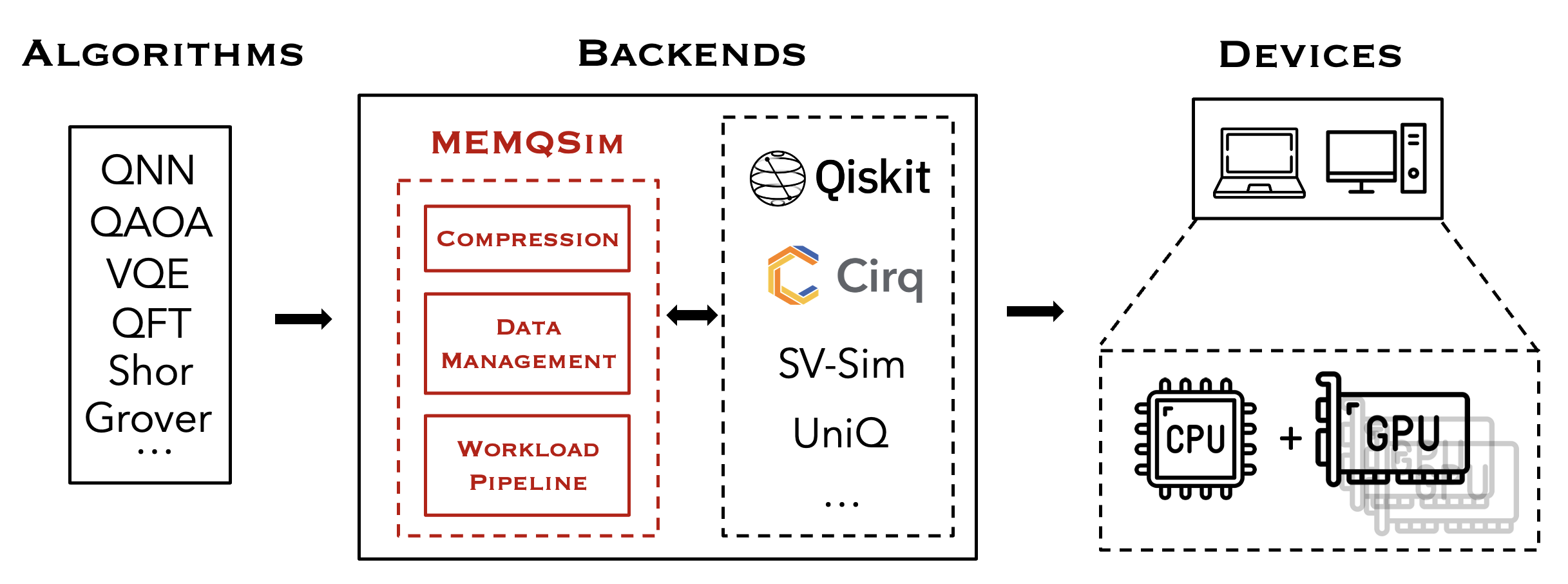}
\vspace{-4mm}
	\caption{Modularized quantum circuit simulation overview.}
	\label{fig:overview}
 \vspace{-8mm}
\end{figure}

In response to these challenges, we introduce a state vector quantum simulator that leverages data compression, termed as \thiswork. The high-level conceptual overview of our approach is depicted in Figure \ref{fig:overview}. Our methodology employs data compression to achieve high memory efficiency, with careful design to ensure efficient simulation. \thiswork{} is independent of quantum algorithm composition and simulation computational tasks, offering significant modularity that allows it to be incorporated into major simulation backends such as Qiskit, Cirq, SV-Sim \cite{li2021sv}, UniQ \cite{zhang2022uniq}, among others. We make the following contributions:

\begin{itemize}
\item We introduce a highly memory-efficient, modular framework. \thiswork{} is designed to be compatible with the integration into state-of-the-art GPU simulator backends.
\item We extend the potential of lossy compression into state-vector simulation to augment the qubit threshold.
\item \thiswork{} effectively manages the massive memory interchange between CPU and GPU. We build a prototype of \thiswork{} and show the preliminary results. 
\end{itemize}

\section{Design and Preliminary Result}
\label{sec:design}

In this section, we present an in-depth design of a memory-efficient state vector simulator via data compression. Our proposed system employs CPU to store the state vector and GPU to execute the state vector updates. This approach is adopted considering that user-level devices typically possess larger CPU memory compared to GPU memory. Yet, the implementation of this design is not straightforward due to the complexities of data management and the intricacies of compression/decompression during the simulation. 

\textit{Design challenges}: (1) The intensive data exchange between the CPU and GPU requires careful scheduling. Since normally GPU memory capacity is much less, the GPU must retrieve data from the CPU, conducting computational operations one partial data at a time. (2) The frequency and granularity of compression and decompression significantly influence the simulation speed. Excessive compression/decompression could result in substantial overhead on the end-to-end time, and a coarser granularity could precipitate a significant memory footprint issue, while excessively fine granularity could lead to a lower compression ratio. (3) Different quantum algorithms' behaviors 
 affect the access pattern on the state vector.

\begin{figure}[h]
\vspace{-4mm}
	\includegraphics[width=.95\linewidth]{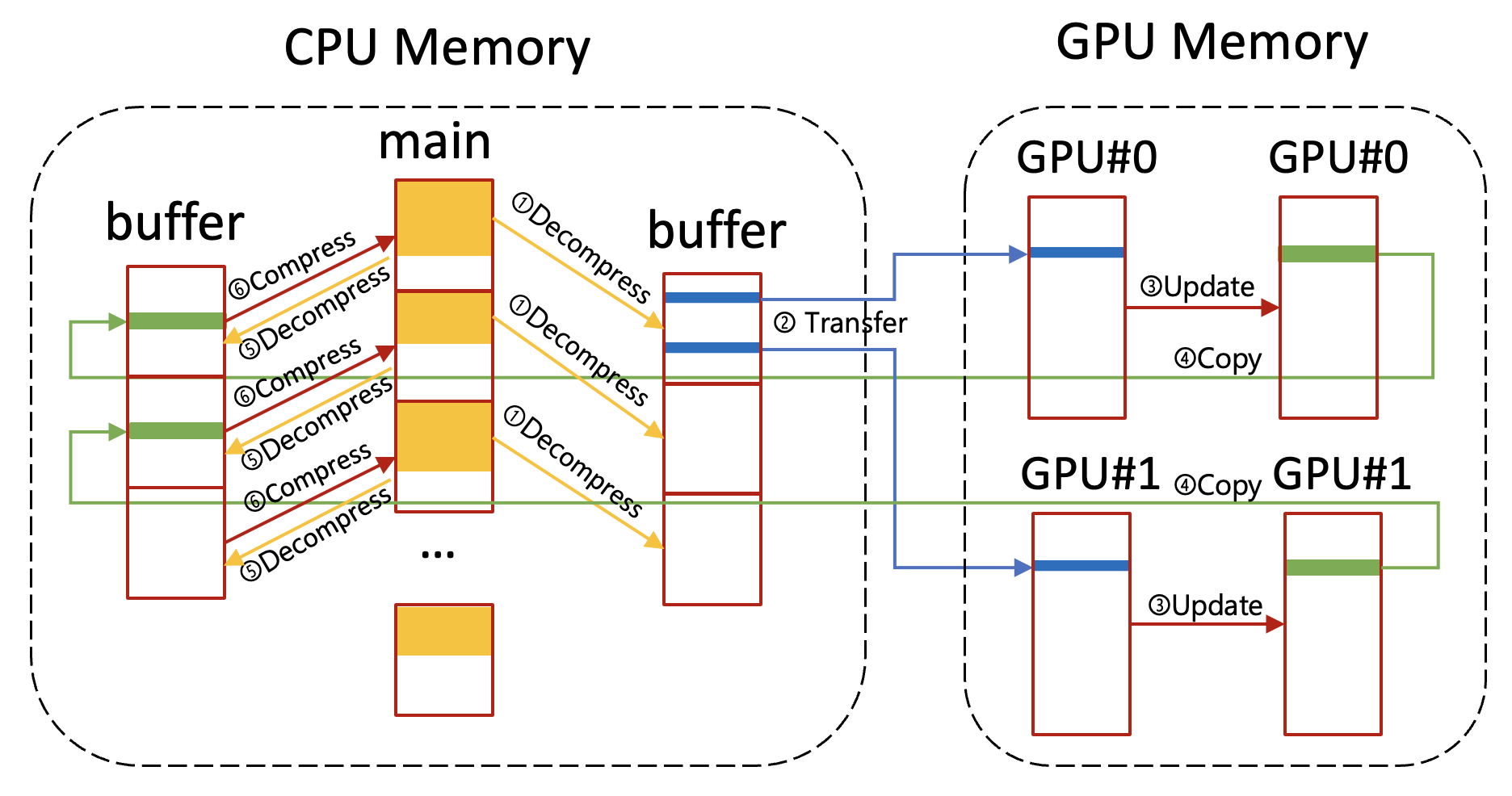}
 \vspace{-4mm}
	\caption{Data management design of \thiswork{}.}
	\label{fig:design}
\vspace{-6mm}
\end{figure}

In light of these identified challenges, we propose our design, \thiswork{}. An overview of our approach is illustrated in Figure \ref{fig:design}. We explain the overall simulation process below:

\textit{Offline stage} 
\thiswork{} partitions the input circuit and the corresponding state vector and each data chunk of the state vector is compressed independently and stored in CPU memory with such compressed format.

\textit{Online stage:} As shown in Figure~\ref{fig:overview}, \thiswork{} pipelines the decompression, buffer transfer between CPU and GPU and GPU computation. In particular, \thiswork{} (1) decompresses a selection of data chunks to the CPU buffers and (2) transfers the corresponding state vector amplitudes to the GPU memory. This process is repeated throughout the entire state vector until the GPU memory is fully occupied with ordered state vector amplitudes. (3) \thiswork{} initiates the GPU kernel asynchronously to update the state vector amplitudes during the CPU-GPU data transfer and (4) returns the updated values back to the CPU buffers. (5) Subsequently, the CPU leverages idle cores to decompress the data chunks and perform updates to the state vector amplitudes on the CPU side. (6) Finally, the data block is re-compressed and stored back into the main memory.
Upon the GPU's completion of a single iteration, the aforementioned procedure is reiterated to update all amplitudes. Subsequently, the process advances to the subsequent stage and continues until all stages have been addressed. We have developed a prototype of \thiswork{}, plugged into the SV-SIM \cite{li2021sv} framework. As we move forward, our design harbours the potential to serve as a plugin for a range of GPU simulators, while also being adaptable to accommodate various compression algorithms.

As of step (2), we have devised two strategies to execute this process. The first approach entails the transfer of corresponding state vector elements to the GPU memory one at a time, utilizing CUDA asynchronous copies. The alternate strategy involves allocating a buffer on the GPU side and shifting the data chunk from the CPU buffer to the GPU buffer. Following this, GPU threads are employed to map all these amplitudes to their appropriate positions.

We present some preliminary results pertaining to the time taken by various data movement strategies between the CPU and GPU, as depicted in \ref{tab:general}. The synchronous strategy entails the transfer of a complete data chunk through a singular CUDA memory copy operation, thereby exemplifying the minimum time necessary for the transfer between the CPU and GPU. As indicated, the host-to-device time associated with the asynchronous strategy is approximately 870 times longer than the synchronous time. This discrepancy is achieved by reducing multiple initiations of CUDA memory copy operations that cause significant overhead. As for the buffer strategy, although it demands additional memory space, it significantly boosts the data movement speed: the time needed for the buffer strategy is only about 1.03x compared to the synchronous version. By employing the state-of-the-art data compressor, we extrapolate that on average 5 more qubits to simulate can be achieved without slowing down the original quantum circuit simulation. 

\begin{table}[t]
\centering
\caption{Data transfer time H2D/D2H in seconds.}
\vspace{-2mm}
\begin{adjustbox}{max width=.9\columnwidth}
 \begin{tabular}{||c | c | c |c ||} 
 \hline
  qubits $\downarrow$& Sync copy time & Async copy time & Buffer copy time \\
  \hline
  20& 0.003/0.008 & 2.7/9.2&0.003/0.004\\
  \hline
  25&0.080/0.233 & 77.9/294.4 &0.110/0.273\\
 \hline
 \end{tabular}
 \end{adjustbox}
 \vspace{-4mm}
 \label{tab:general}
\end{table}

\section{Conclusion and Future Work}
In this extended abstract, we have introduced a highly memory-efficient state vector simulation of quantum circuits premised on data compression, harnessing the capabilities of both CPUs and GPUs. We have elucidated the inherent challenges in architecting this system, while concurrently proposing our tailored solutions. Moreover, we have delineated our preliminary implementation and deliberated upon the potential for integration with other GPU-oriented simulators. In forthcoming research, we aim to present a more comprehensive set of results, bolstering the assertion of the efficacy and performance of our approach.

\bibliographystyle{plain}
\bibliography{refs}

\end{document}